# Metaverse: Security and Privacy Concerns

Ruoyu Zhao, Yushu Zhang, Youwen Zhu, Rushi Lan, and Zhongyun Hua

***Abstract*—The term "metaverse", a three-dimensional virtual universe similar to the real realm, has always been full of imagination since it was put forward in the 1990s. Recently, it is possible to realize the metaverse with the continuous emergence and progress of various technologies, and thus it has attracted extensive attention again. It may bring a lot of benefits to human society such as reducing discrimination, eliminating individual differences, and socializing. However, everything has security and privacy concerns, which is no exception for the metaverse. In this article, we firstly analyze the concept of the metaverse and propose that it is a super virtual-reality (VR) ecosystem compared with other VR technologies. Then, we carefully analyze and elaborate on possible security and privacy concerns from four perspectives: user information, communication, scenario, and goods, and immediately, the potential solutions are correspondingly put forward. Meanwhile, we propose the need to take advantage of the new buckets effect to comprehensively address security and privacy concerns from a philosophical perspective, which hopefully will bring some progress to the metaverse community.**

## I. Introduction

CAN Alice engage in immersive interaction with her friends who live thousands of miles away? Can Bob seamlessly move from the movie theater to the shopping center in an instant? Can Peter who has a leg disability stand and run like a normal person? Many people around the world may have such similar questions every day.

A recently popular term, metaverse, may be able to address these questions easily, and in fact, it is not a newborn but a palingenesis. The term "metaverse" was first invented in a novel named *Snow Crash* in 1992 [1], which was a combination of two words, "meta" and "verse". The former means beyond reality, i.e., in a virtual environment; the latter refers to the universe, which means that people can immerse themselves in this environment for living like reality. Since the term was put forward, its definition has been very diverse [2], e.g., lifelogging, future social networks, next generation Internet, and virtual world, which makes it cast a layer of mystery. Overall, however, a consensus is that users living in the real world link to and operate their avatars who are in the metaverse through access terminals in order to immerse themselves into a three-dimensional (3D) virtual world as shown in Fig. 1.

In a nutshell, various information including instructions of users is collected by sensors to sent the terminal; the terminal synthesizes the information from sensors and then sends it to the server by Internet to control the corresponding avatar in the metaverse; servers comprehensively process the information of a large number of users and reflect it in the 3D virtual world; each legal avatar activity sent by users is executed. Conversely, the status in the metaverse when avatars execute activities is fed back to servers, e.g., if a work of art is created by an avatar, its information such as content needs to be told to servers; servers record and generate the corresponding 3D virtual scenario that is broadcasted all user terminals; terminals display the real-time scenario in the metaverse to users after receiving server information and further send detailed instructions to sensors; each sensor sends a corresponding signal according to the instruction to immerse users into it. For example, it will be more realistic for the user if the sensor on the hand responds appropriately when the corresponding avatar shakes hands with others in the metaverse.

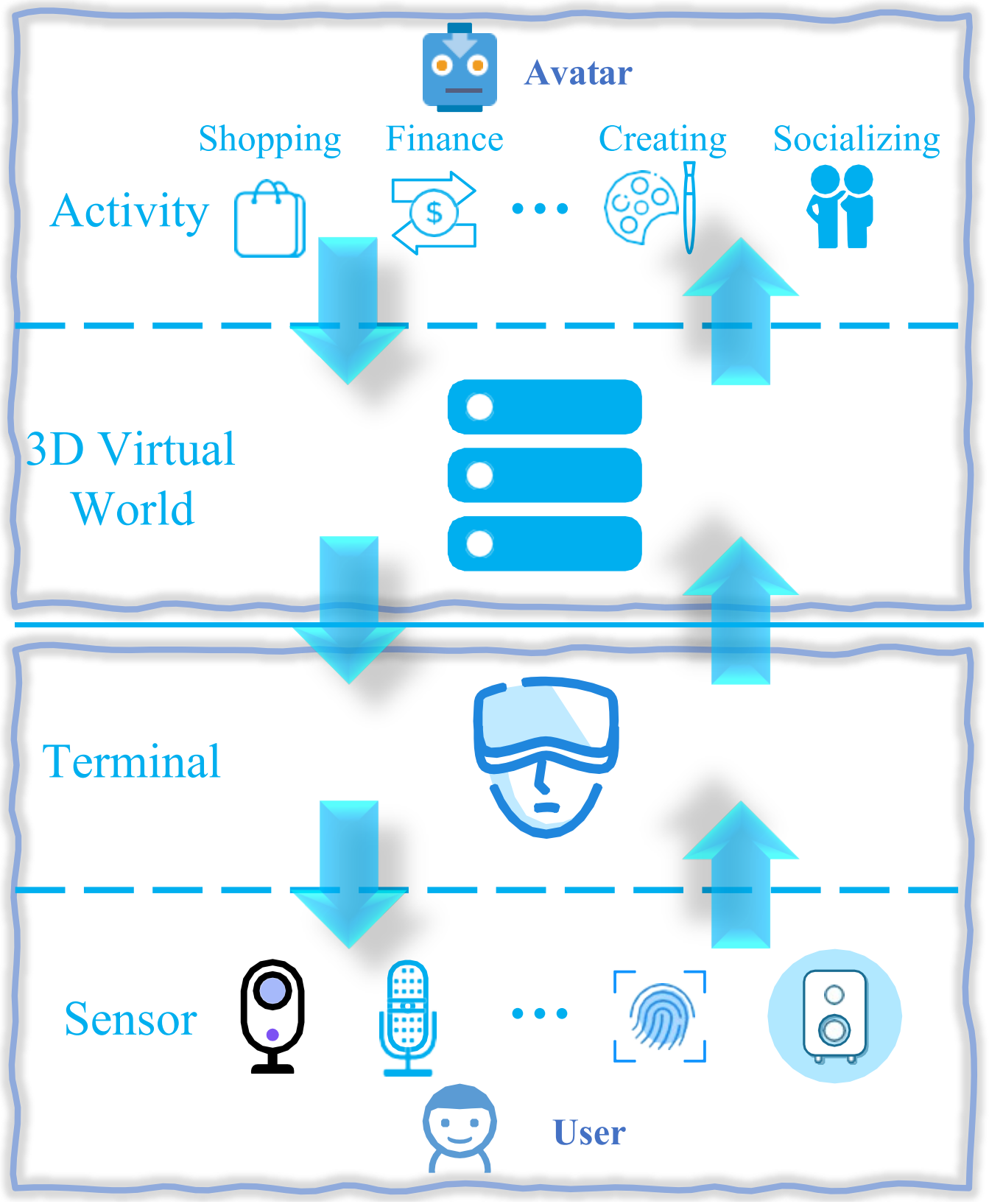

Fig. 1. The basic infrastructure of the metaverse.

The metaverse can bring a lot of benefits to people in the real universe. The problems mentioned at the beginning of this article can be well addressed in the metaverse owing to the characteristic of virtuality. Meanwhile, the long-standing problem of discrimination in the reality may be alleviated.

R. Zhao, Y. Zhang, and Y. Zhu are with the College of Computer Science and Technology, Nanjing University of Aeronautics and Astronautics, Nanjing 211106, China (e-mail: zhaoruoyu@nuaa.edu.cn; yushu@nuaa.edu.cn; zhuyw@nuaa.edu.cn).

R. Lan is with Guangxi Key Laboratory of Image and Graphic Intelligent Processing, Guilin University of Electronic Technology, Guilin 541004, China (e-mail: rslan2016@163.com).

Z. Hua is with the School of Computer Science and Technology, Harbin Institute of Technology (Shenzhen), Shenzhen 518055, China (e-mail: huazhongyun@hit.edu.cn).

For instance, people with physical disabilities can move like ordinary people in the metaverse as long as they are conscious; there is no difference in physical strength between the elderly and the young; gender is no longer innate; looks can change at will; and skin color and race are no longer have to be known to others.

On the other hand, the metaverse is confronted with new and serious security and privacy concerns despite its obvious value. First, more important and sensitive information in the real world through terminals may be stolen by malicious others since avatars have a closer relationship with users than other virtual worlds such as online games. Second, avatars have a lot of interaction with other avatars and non-player characters, which are not all intended to be understood by others. Third, there will inevitably be scenarios in the universe that make some people feel inappropriate due to the differences of culture and others, not even excepting the metaverse, and not to mention malicious avatar behaviors such as harassment. Fourth, the ownership, illegal copy, and transaction of goods in the metaverse are also thorny challenges. The simplest way to solve the security and privacy concerns of the metaverse is to prohibit users from entering it [3], but this crudest method completely gives up the its benefits that is just throwing out the baby with the bath water. In this article, we focus on the potential emerging security and privacy concerns of the metaverse itself and then propose alternative solutions that do not completely damage the interests. The key points of this article can be summarized as follows:

- We analyze the concept of the metaverse and propose that it is a super 3D virtual-reality (VR) ecosystem compared with other VR technologies.
- The serious challenges of security and privacy concerns in the metaverse are pointed out and summarized.
- Some potential solutions for these security and privacy concerns in the metaverse are proposed correspondingly.
- The new buckets effect is applied to think philosophically about how to deal with security and privacy concerns in a comprehensive way in the metaverse.

## II. Overview of Metaverse

Intuitively, the boundary between the metaverse and VR, augmented reality (AR), and mixed reality (MR) appears to be hazy. In fact, the metaverse can be highly summarized as a super virtual-reality *ecosystem* based on the Internet, which is composed of inter-disciplinary technologies as shown in Fig. 2, e.g., VR, AR, MR, artificial intelligence, machine learning, computer vision, speech recognition, blockchain, and the Internet of things. By contrast, VR/AR/MR is only a kind of virtualized and digitized technology, and it does not necessitate a comprehensive ecosystem, rules, and the Internet, despite being an important component of the metaverse.

The term "ecosystem" implies that the components of the metaverse interact and restrict each other, and are in a relative stable dynamic equilibrium state, forming a persistent and unified virtual world. Meanwhile, a large number of users are the foundation of the ecosystem. If there are no users, it can only be labeled a 3D virtual vision system rather than be called

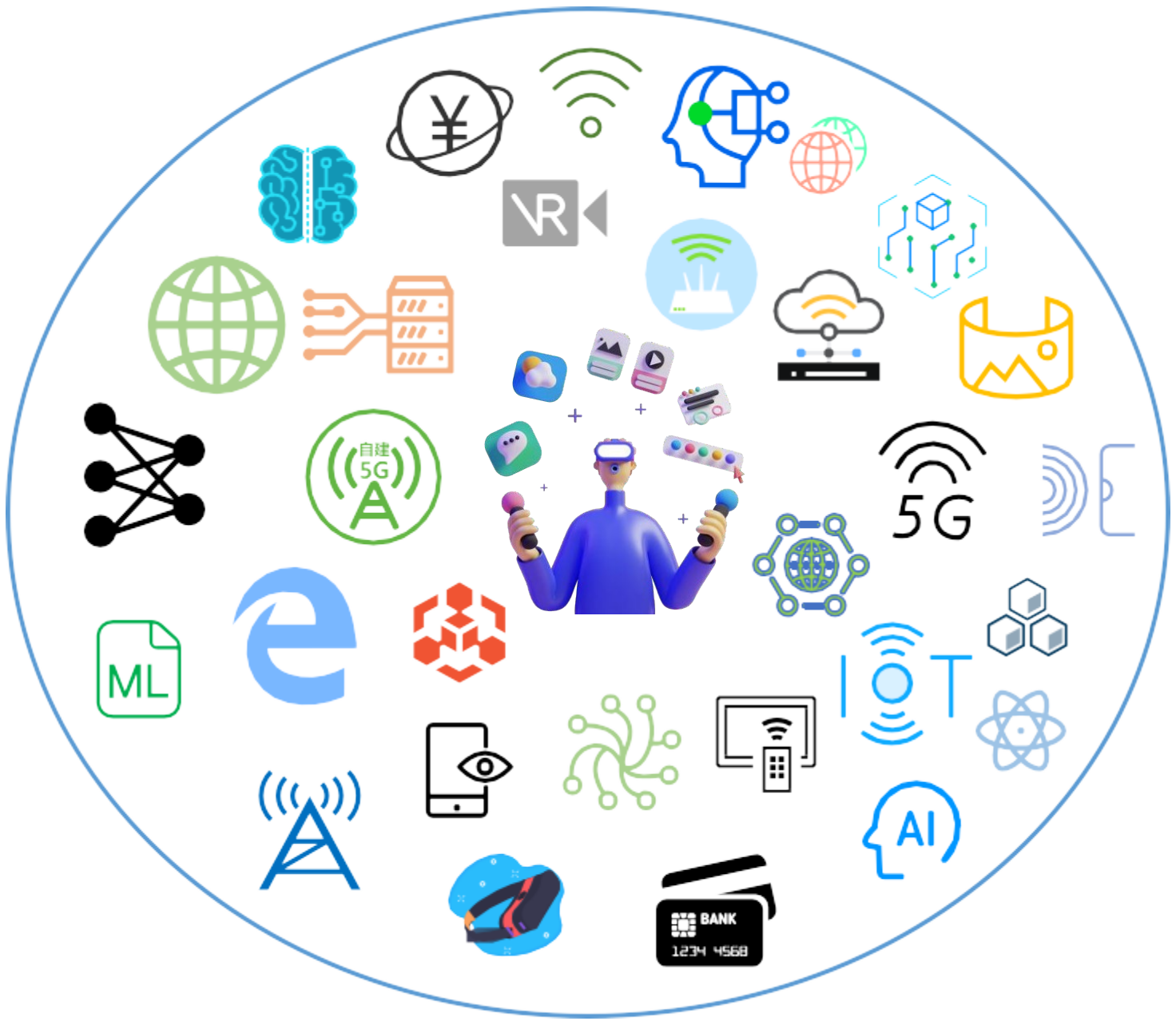

Fig. 2. An illustration of the main technology composition of the metaverse.

"verse" no matter how perfect it is. Just like a place with all kinds of goods but without customers who pay the bill, it can only be called a warehouse rather than a shopping mall. In truth, users create demand to stimulate the development of the metaverse, which in turn attracts users to enter, resulting in a positive ecosystem. In other words, a metaverse without users is doomed to failure, which also implies that perhaps only a few metaverse platforms will eventually flourish and the others will die. This trend is already evident on current Internet platforms, e.g., people prefer to choose Instagram for sharing pictures and Tiktok for short videos, despite the availability of alternative ones.

There are two main reasons why the metaverse recently can be palingenesis after this concept was put forward many years. First, the COVID-19 epidemic has trained people to be familiar with the virtual digital world and promoted the socializing to shift from offline to online to some extent [4]. Second, the recent significant progress like Big Bang of the above related technologies as shown in Fig. 2 makes it possible to build a metaverse technically.

## III. Security and Privacy Concerns

The development of anything is inevitably accompanied by security and privacy concerns with no exception to the metaverse. Specifically, these concerns can be divided into four categories:

- User information: multi-sensor fusion is one of the characteristic of the metaverse as shown in Fig. 1, making a large amount of user information to be collected. There is no doubt that sensors are necessary since they help users to improve the experience resulting in immersing themselves in the metaverse. On the other hand, although many users may not be noticed or even realized the problem [3], some user information collected by sensors, e.g., related physiological, physical, biometric, and social, is too personal. If it is leaked, it will greatly endanger the security and privacy of users [5]. Hence, it is critical that user information is protected.

- Communication: one of the features of the metaverse is its high interactivity and sociality, and thus a lot of communication inevitably takes place. Many activities, e.g., sharing, cooperating, and increasing mutual trust and understanding, in the metaverse are difficult to be done without the help of communication. Although it may not contain the above-mentioned user information, most users are nevertheless unwilling to tell those who are non-communicators since communication content is highly private and sensitive. As a result, it is important to protect communication and it should be done in such a way that non-communicators are prevented from comprehending and recovering the contents of the communication while legal communicators can.
- Scenario: it is conceivable to encounter the same security and privacy concerns as the real realm since the metaverse is a surreal universe. There are two main aspects to be considered: the scenario itself and avatars in the scenario. For the former, as a great number of users are clustered on a metaverse platform (and in fact there are not many alternative platforms to pick from), their understanding of cultures, religions, and so on will inevitably vary. Therefore, the scenario will not meet everyone's wish and even cause misunderstanding for some avatars. For the latter, the influx of users will inevitably introduce some malicious and immoral ones who may insult, track, or even sexually harass other avatars in the metaverse, and these activities have appeared in online games [6].
- Goods: the metaserve has the characteristics of imagination, high creativity, high degree of degree of freedom, and high personalization. Thus, avatars can create all kinds of goods, such as the character modeling, appearance, costumes, buildings, and artworks, according to personal wishes. These goods can be applied by creators or sold, i.e., they are either created through efforts or at the cost of money (of course, they may also be freely given by friends), implying that they include both spiritual and financial values. Avatars do not wish the value to be illegally damaged, e.g., an avatar tailors a personalized dress for themselves and may not want to see it on others. Meanwhile, goods transactions may also be damaged by malicious users and avatars also have a demand to anonymize rights in the transactions. Hence, it is important for the secure protection of goods themselves and transactions in the metaserve.

## IV. User Information

User information has always been a very important and sensitive concern related to security and privacy in modern society. More detailed user information in the metaverse will be collected than previous platforms such as society networks due to the characteristics of the immersion, the indistinguishability of virtual and reality, and multi-sensor. This makes illegal third parties more interested in this information and may attack through the network as shown in Fig. 3. Meanwhile, the information is no longer directly under the actual physical control of users as long as it is transmitted out from the terminal, which implies that information is at risk after leaving the terminal, and in other words, the protection of user information needs to be carried out at the terminal.

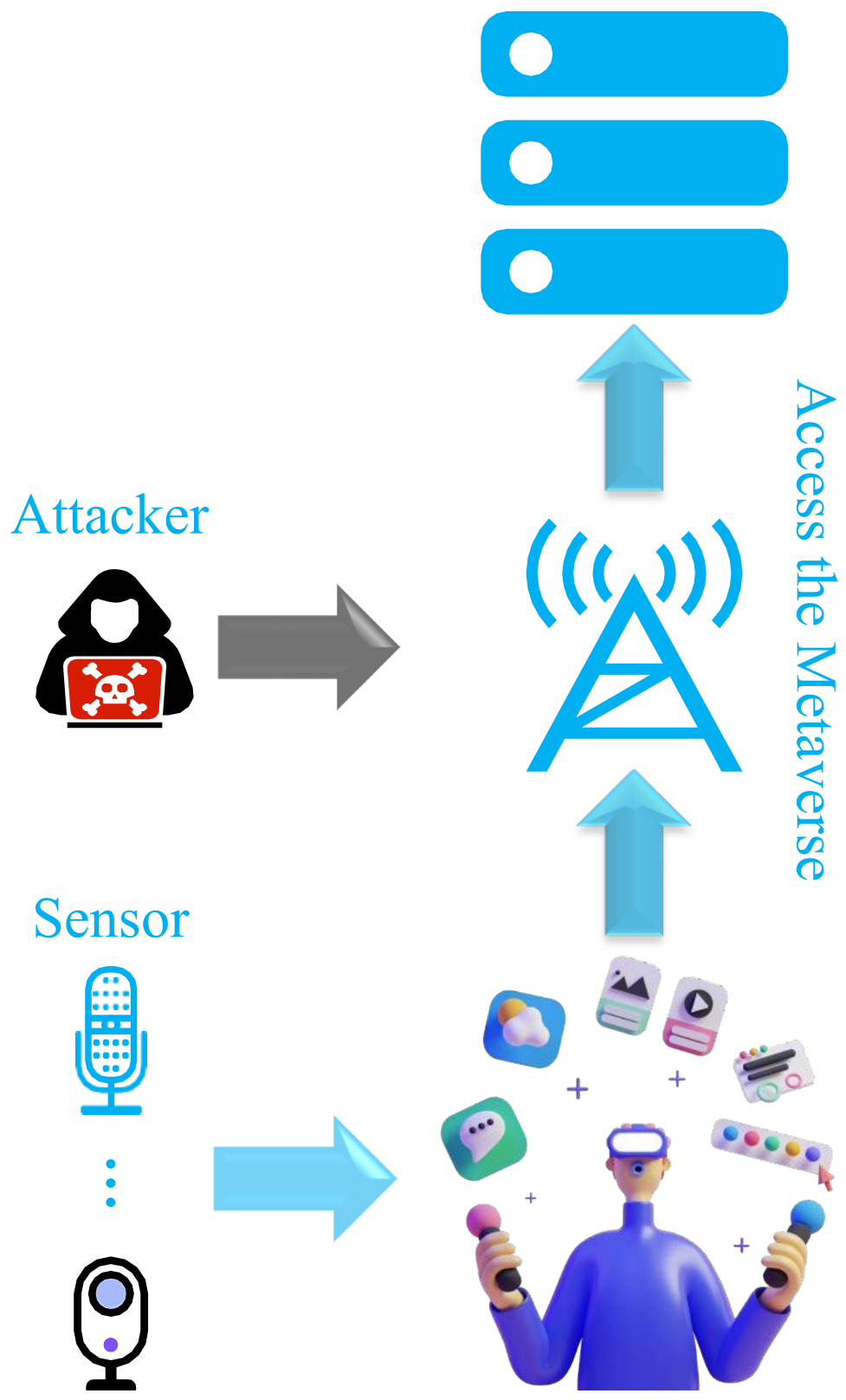

Fig. 3. An illustration of the metaverse being attacked by the attacter.

It is worth noting that everyone has vary views on privacy since everyone has different cultural habits and acceptance. In addition, any solution cannot protect all information without paying any price. Therefore, the solution should be targeted protection for the goal that users want to achieve, and no one can perfectly prevent from all risks except to give up the use directly. Next, we will state some solutions to protect user information in the metaverse.

For single accurate signal information such as heartbeat information obtained by the sensor, it can be protected by only shielding the signal and prohibiting transmission. On the other hand, visual multimedia including image and video, which occupy the mainstream in the metaverse, contains a lot of sensitive information and much accurate information can also be extracted from it, e.g., heart rate [7], health, and social status. It can not be simply shielded since visual multimedia is inextricably related to the application of the metaverse. Therefore, specific treatment and protection are needed, which can be classified into three categories: generalized, white-list, and black-list.

*Generalized protection* means that users desire to protect some visual content since they feel there is probably a privacy issue. It is protected in a general way without considering specific and fine privacy compared with the solutions of white-list and block-list. In the multimedia visual content, only one part may be needed and other parts are redundant. For example, it only needs people to appear and background which may reveal too much information is unnecessary in the video conferencing in the metaverse. This problem can be solved by matting, i.e., the visual content can be divided into those that need to be preserved and abandoned and then processed accordingly. It has been applied in practice, e.g., Zoom and Tencent Conference have allowed users to choose virtual background options a year ago. Similarly, the solutions such as face swapping and 3D model replacement can be applied in light of the privacy risk that faces may pose.

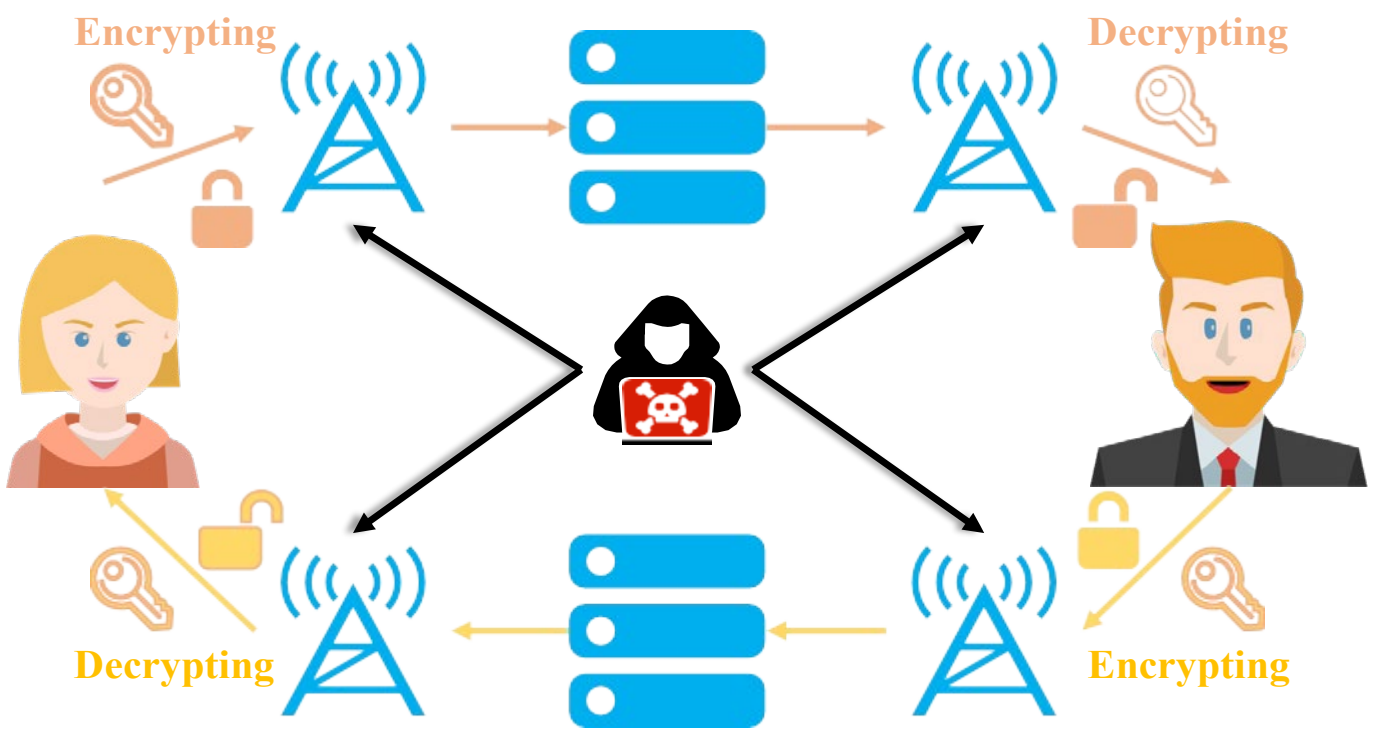

Fig. 4. An illustration of the communication being protected.

*White-list protection* refers to that everything is processed and protected in addition to the information selected by users (which similar to the white list), and this is a targeted protection compared with the above. As a simple example, a smile competition is organized in the metaverse and the avatar with the brightest smile can win the game. The face content of the user is required for participating in the competition, but the user may only wish to employ the facial content for analysis smile. Hence, the face in the multimedia should not contain any information except smile. For this problem, Wu *et al.* proposed a solution to train a model through machine learning to protect visual content, which only retains the usability of specific information but other will be deleted and cannot be extracted [8].

*Black-list protection* implies that nothing about the visual content in the multimedia is processed except what users choose (which similar to the black list). This kind of protection often aims at the face, resulting in a highly accurate with good visual observability. For such protection, faces are often destructed into specific vectors for further accurate processing and each vector represents a signal. Such vectors are typically separated into two classes: identity and attributes. Some vectors pertaining to the information that users desire to protect are processed to be protected while the rest remain unchanged. Then, these vectors are integrated and reconstructed into the protected face by generating models called anonymization and attribute protection, respectively, depending on whether the identity or attribute is processed.

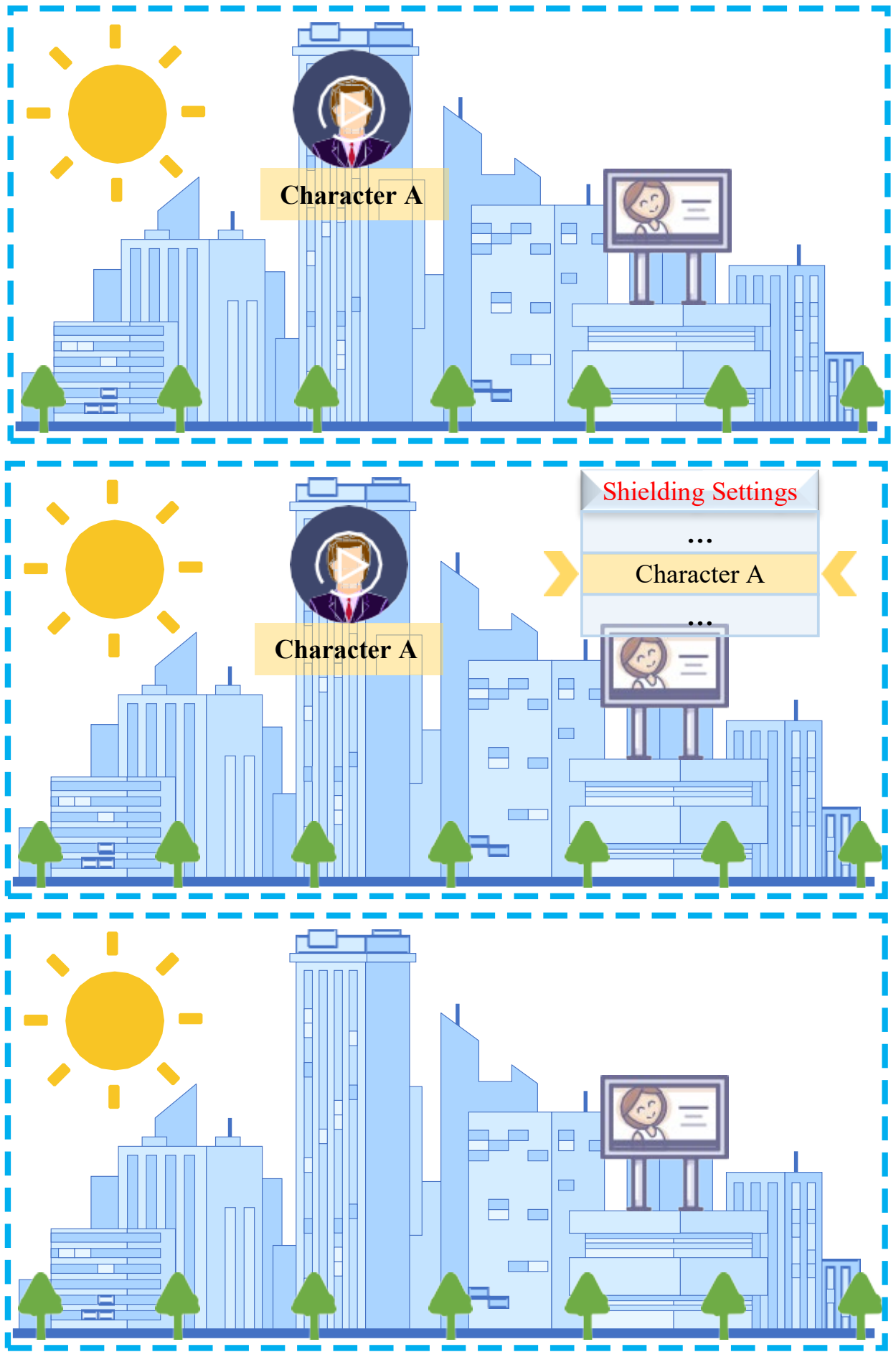

Fig. 5. An example of the scenario shielded according to user selection.

## V. Communication

Interaction and social are required in the metaverse and thus frequent communication occurs. Communication is carried out by avatars on the surface but it is actually controlled by users. For the participants in the communication, they only want the target party, i.e., legitimate receiver, to know the content of the communication, while the third party is unaware. It indicates that the content can not be directly eliminated in the solution, as in the case of user information, but it must be able to be recovered for the target party.

A powerful solution to this goal is encryption as shown in Fig. 4, i.e., the sender sends the information after encrypting with the key and the legitimate receiver decrypts it employing the correct key after receiving it. Thus, the meaningless ciphertext is transmitted in the course of communication and attackers cannot decrypt it as long as they do not have the correct key even if the ciphertext is intercepted. By the way, the key and encryption algorithm are primarily responsible for the security of the ciphertext, i.e., the ability to prevent attackers without the correct key from breaking the ciphertext. For the key, it should be long enough, i.e., the key space is large enough, to prevent the attacker from using exhaustive attack, and meanwhile, the communication will be in danger once the key is leaked and thus the key should be kept confidential and replaced on a regular basis. For the encryption algorithm, it should go through enough cryptographic tests and analysis to prevent possible attacks, and then the algorithm that meets the corresponding security standards is selected according to the actual needs.

Although general encryption schemes for visual multimedia including video and image, converting the meaningful visual content into useless noise-like one, have been able to meet the needs of security and privacy, they do not think about visual obeservability. In fact, this feature may be important for enjoying the communication. Senders share a large number of encrypted images that are all noise-like ones and cannot be distinguished by browsing, and in other words, they cannot be selected unless all of them are decrypted, which may be a poor experience for receivers. Thumbnail-preserving encryption can be applied to alleviate the contradiction between privacy security and visual observability [9], which erases the fine details of the visual content but preserves the coarse one. The final result is similar to the mosaic effect, which implies that certain visual information can be obtained by browsing the ciphertext but the details cannot be learned until decrypting.

## VI. Scenario

Some conflicts arise from time to time in the real world, such as those arising from religious, political, gender, and sexual minorities, as a results of various cultures and ideas from different people. This phenomenon is even more severe in the virtual world due to lack of distance and other restrictions, e.g., cyberbullying on social networks and scale violence in online games, which will undoubtedly degrade the experience and create discomfort. But for users of current network platforms, they can escape from these unpleasant places and form small virtual communities with people who have similar interests and opinions so as to avoid some negative effects. However, this way does not seem to exist for the metaverse since itself is a complete universe and it is difficult for a tiny group of individuals to make a fresh start. Meanwhile, just as in the real world, there are malicious avatars in the metaverse who are evil to others, such as harassment and stalking [3].

For a user-friendly metaverse environment, a setting window should be provided to allow users to set to prevent some scenarios around their avatars. For example, a propaganda video of a political figure is playing outside a building, but some avatars dislike him/her and thus they choose to shield him out as shown in Fig. 5. It is worth noting that the scenario does not disappear from the metaverse but can not be seen for the specific avatar, which can also be called as the personalized scenario presentation. Similarly, for the offensive and insulting content of other avatars' speeches and texts, specific keywords can also be set to detect them through the voice and text detection model for shielding.

On the other hand, this solution is difficult to detect avatar aggressive and bullying behavior. First, the meaning of avatar behaviors is often subtle and whether it is malicious needs to be combined with the actual situation and context. Second, behaviors themselves are highly diverse and many malicious ones have no clear definition compared with the speech and text. For example, it is not always deemed malicious for an avatar to play with a gun, but it is malicious if the gun points towards other avatars, making it is impossible to judge whether there is malicious by simply detecting the gun unless supplemented with specific circumstances. According to researches, the detection of malicious bullying by the avatar can combine multiple factors such as body-pose, facial emotion, hand gesture, object, and social, resulting in a satisfactory outcome [10].

For the harassment and stalking, the preceding solution is not useful since even though these avatars are shielded in our own scenario, we still exist in malicious avatar scenarios and they can still do this kind of activity. A good solution for it is to disappear suddenly, e.g., cloaking and teleportation [3], and as a result, the malicious avatars cannot find the target one. Furthermore, a user can create multiple avatars and randomly select different ones each time when he/she accesses the metaverse to prevent some ill-intentioned avatars from looking for patterns over time.

## VII. Goods

As Lee *et al.* pointed out [2], creation is an important part of the sustainable development of the metaverse. In addition, the high degree of freedom and open environment of the metaverse also greatly encourages the emergence of the creation activity driven by individualized psychological needs or money, which means that a large number of goods produced by creation will appear. No matter what the motivation, the owner of the goods will not want others to illegally copy and abuse it. Based on this point, it is necessary to take methods to protect it.

A viable solution is invisible watermarking, a technique aimed at embedding a specific mark related to identity in the goods as they are created or the ownership is transferred. It will not affect the visual effect of its own goods in the metaverse owing to invisibility, and can be extracted or detected when needed. Therefore, some functions, including content protection, authentication, and tamper-resistance, are realized [11], which further deters malicious avatars from stealing and illegally copying goods. Moreover, compared with the real-world scenario, watermarking is more suitable for the metaverse. In the real world, it is essentially a modification of visual content no matter how invisible it is, which will destroy some physical features, and thus the existence of watermarking can be illegally detected by some technical means. This difficulty, however, does not exist in the metaverse which is a generative digital scenario rather than a physical one, and it is difficult to detect it by existing means even if the watermarking is made on it.

The blockchain is an excellent solution to the problems of ownership, traceability, and transfer of goods, which has the following characteristics: decentralization, tamper resistant, and anonymity [12]. Decentralization enables each avatar to participate in blockchain activities fairly, leading that avatars are able to register the ownership of each goods by themselves, which is the premise of protection. Tamper resistant is due to the fact that tampering with the blockchain require more than 51% of the computing power support in the system. It is not the interests of those with a lot of computing power since they need to maintain the stability of the system to obtain the maximum benefits. Therefore, there is no need to be concerned about its effectiveness following ownership registration. Anonymity allows the avatar not to worry about disclosing who has the ownership after registration, nor about the exposure of its identity during a transaction. Furthermore, the blockchain has an effective tool, the smart contract [13], which can prevent both parties from defaulting in the transaction of goods. Specifically, the two sides of the transaction need to reach an agreement in advance and then draw up a contract to sign. Once the conditions stipulated in the contract are met, it begins to be executed automatically and is not changed by human will. Hence, the transaction of the ownership of goods in the metaverse can be carried out safely.

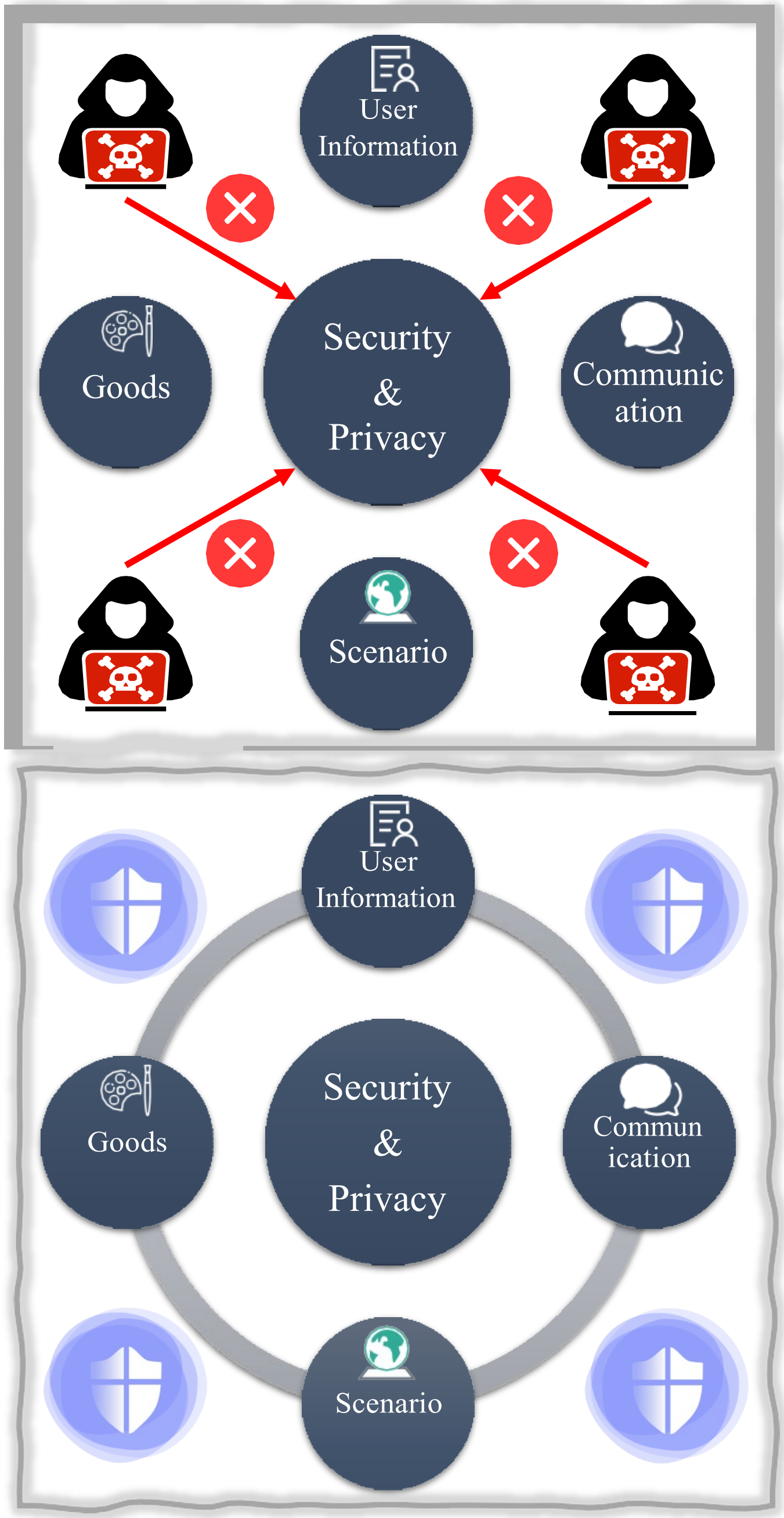

Fig. 6. An example of the new buckets effect for protecting security and privacy.

## VIII. New Buckets Effect

There is no doubt that the above solutions can achieve a protective effect in dealing with their corresponding security and privacy concerns when standing at the angle of the raised problem. Meanwhile, it is common for researchers to notice the shortcoming of existing solutions and strive to continuously improve them to achieve better effects, like classic buckets effect, in which each board can be enlarged to contain more water. However, there are gaps between each board, and if it is not fastened, no matter how much water is contained, it will leak. Similarly, if people only consider a single problem in isolation and turn a blind eye to others, they may be able to put forward corresponding and effect solutions, but they are not very helpful to mitigate the security and privacy of the whole ecosystem as shown in the top panel of Fig. 6. For example, when an avatar chat with others in the mataverse, he/she may disclose trivial daily in unexpected ways, and then expose his/her user information from the real world. If the intentional avatar takes advantage of this and deliberately communicates with the avatar, the user information may be successfully obtained in this way.

Therefore, it is necessary to add a lock ring on the bucket to minimize water leakage, i.e., new buckets effect. Similarly, a comprehensive rather than peer-to-peer consideration of how to address security and privacy concerns, coming up a package of solutions, is necessary. It should be to allow users to choose and execute multiple and closely related solutions simultaneously, which may effectively alleviate concerns in the metaverse as shown in the bottom panel of Fig. 6. The statement may help better protect the security and privacy of the metaverse, but it also poses a greater challenge for researchers, which makes them need to design a systematic and coherent solutions based on the thinking of comprehensive and global.

## IX. Conclusion

Security and privacy concerns are inevitable in the development of everything and need to be solved. In this article, the concept of the metaverse is first refined and the security and privacy concerns of the metaverse are analyzed and summarized involving user information, communication, scenario, and goods. Then, through reflecting on the key problems we summarized, corresponding potential solutions based on shielded, machine learning, encryption, watermarking, blockchain, and so on are proposed, which adhere to the user-centered, allowing users to choose personalized and appropriate ones to address these concerns. Last, through philosophical reflection, this article puts forward the idea of drawing lessons from the new barrel effect for comprehensively and effectively alleviating the security and privacy concerns of the ecosystem. We hope that the relevant concerns we raised can attract attention in the metaverse community and provide some assistance in mitigating them.

## References

[1] N. Stephenson, *Snow crash: A novel*. Spectra, 2003.

[2] L.-H. Lee, T. Braud, P. Zhou, L. Wang, D. Xu, Z. Lin, A. Kumar, C. Bermejo, and P. Hui, "All one needs to know about metaverse: A complete survey on technological singularity, virtual ecosystem, and research agenda," *arXiv preprint arXiv:2110.05352*, 2021.

[3] B. Falchuk, S. Loeb, and R. Neff, "The social metaverse: Battle for privacy," *IEEE Technol. Soc. Mag.*, vol. 37, no. 2, pp. 52–61, 2018.

[4] Z. Chen, H. Cao, Y. Deng, X. Gao, J. Piao, F. Xu, Y. Zhang, and Y. Li, "Learning from home: A mixed-methods analysis of live streaming based remote education experience in chinese colleges during the covid-19 pandemic," in *Conf. Hum. Fact. Comput. Syst. Proc.*, ser. CHI '21, 2021.

[5] H. Liu, X. Yao, T. Yang, and H. Ning, "Cooperative privacy preservation for wearable devices in hybrid computing-based smart health," *IEEE Internet Things J.*, vol. 6, no. 2, pp. 1352–1362, 2019.

[6] J. Fox, M. Gilbert, and W. Y. Tang, "Player experiences in a massively multiplayer online game: A diary study of performance, motivation, and social interaction," *New Media Soc.*, vol. 20, no. 11, pp. 4056–4073, 2018.

[7] Q. Zhu, M. Chen, C.-W. Wong, and M. Wu, "Adaptive multi-trace carving for robust frequency tracking in forensic applications," *IEEE Trans. Inf. Forensic Secur.*, vol. 16, pp. 1174–1189, 2021.

[8] H. Wu, X. Tian, Y. Gong, X. Su, M. Li, and F. Xu, "DAPter: Preventing user data abuse in deep learning inference services," in *Proc. World Wide Web Conf.*, 2021, pp. 1017–1028.

[9] Y. Zhang, R. Zhao, X. Xiao, R. Lan, Z. Liu, and X. Zhang, "HF-TPE: High-fidelity thumbnail-preserving encryption," *IEEE Trans. Circuits Syst. Video Technol., in press, doi: 10.1109/TCSVT.2021.3070348*, 2021.

[10] N. Vishwamitra, H. Hu, F. Luo, and L. Cheng, "Towards understanding and detecting cyberbullying in real-world images," in *IEEE Int. Conf. Mach. Learn. Appl.*, 2021.

[11] M. Begum and M. S. Uddin, "Digital image watermarking techniques: A review," *Information*, vol. 11, no. 2, p. 110, 2020.

[12] H.-N. Dai, Z. Zheng, and Y. Zhang, "Blockchain for internet of things: A survey," *IEEE Internet Things J.*, vol. 6, no. 5, pp. 8076–8094, 2019.

[13] S. Lee, M. Kim, J. Lee, R.-H. Hsu, and T. Q. S. Quek, "Is blockchain suitable for data freshness? an age-of-information perspective," *IEEE Netw.*, vol. 35, no. 2, pp. 96–103, 2021.